\documentclass[a4paper]{jpconf}
\usepackage{graphicx}
\usepackage{amsmath}
\usepackage{amssymb}
\usepackage{ulem}
\usepackage{mathrsfs, mathtools, mhsetup,slashed}
\usepackage{epstopdf}

\def\ket#1{\mathinner{|{#1}\rangle}}
\def\braket#1{\mathinner{\langle{#1}\rangle}}
\newcommand{\pim}{\pi^{-}}
\newcommand{\pip}{\pi^{+}}
\newcommand{\antip}{\bar{p}}
\newcommand{\process}{ p ~\antip \rightarrow \pip ~ \pim}
\newcommand{\gev}{~\mbox{GeV}^2}

\begin{document}

 \title{First results of proton antiproton annihilation into a pion pair at large scattering angles within the handbag approach}
 \author{A T Goritschnig$^1$, S Kofler$^2$ and W Schweiger$^3$}
 
 \address{Institute for Physics, Karl-Franzens-University Graz, Universit\"atsplatz \( 5 \), \( 8010 \) Graz, Austria} 
 
 \ead{$^1$alexander.goritschnig@uni-graz.at}
 \ead{$^2$stefan.kofler@uni-graz.at}
 \ead{$^3$wolfgang.schweiger@uni-graz.at}

 \begin{abstract}
We propose to describe the process \( \process  \) in a perturbative, QCD motivated framework in which a hard $ud \, \bar{u} \bar{d} \rightarrow d \, \bar{d} $ annihilation factorizes from soft transition distribution amplitudes.
We advocate that the scale allowing for this factorization is the large transverse momentum transfer.
In our simplified model, in which the proton is considered as a (scalar)diqark-quark system, a transition distribution amplitude describes the non-perturbative transition of the proton to the meson by emission of  a scalar, isoscalar \( ud \)-diquark and absorption of an antiquark (analogously for $ \bar{p} \rightarrow \pi^- $).
We model the transition distribution amplitudes as an overlap of light-cone wave functions and present first results for the differential cross section.
This process will be measured by the \(\bar{\mbox{P}}\)ANDA experiment at GSI-FAIR.
\end{abstract}

 \section{Introduction}
 The \(\bar{\mbox{P}}\)ANDA detector at FAIR (Facility for Antiproton and Ion Research) in Darmstadt is ideally suited for the study of exclusive channels in proton-antiproton collisions.
 Studying these reactions will open a new area of looking into the interior of hadrons and thus bring new perspectives to the question on how hadrons can be understood in terms of the fundamental degrees of freedom of QCD, i.e. in terms of quarks and gluons.
 One possibility is to investigate the process \( \process \). 
 Pion production is particularly interesting and illuminating because one has well established constraints at hand to find realistic light-cone wave functions (LCWFs) which serve as model input.
 We adapt the description of heavy \( D\)-meson production within the handbag approach, cf. Ref.~\cite{gor1}, to the case of light meson production.
 We advocate that for large transverse momentum transfer and for restricted parton virtualities and intrinsic transverse momenta the process amplitude factorizes into a (hard) partonic subprocess and (soft) hadronic transition matrix elements.
 The hadronic transition matrix elements can be parameterized by transition distribution amplitudes (TDAs) which have been introduced in a series of papers, cf. Refs.~\cite{tda1}-\cite{tda4}.
 To model the TDAs we extend the formalism of LCWF-overlap, developed in Ref.~\cite{kroll2}.

 \section{The Handbag Mechanism}\label{sec_handbag}
 \begin{figure}[h]
\centering
\includegraphics[width=.45\textwidth]{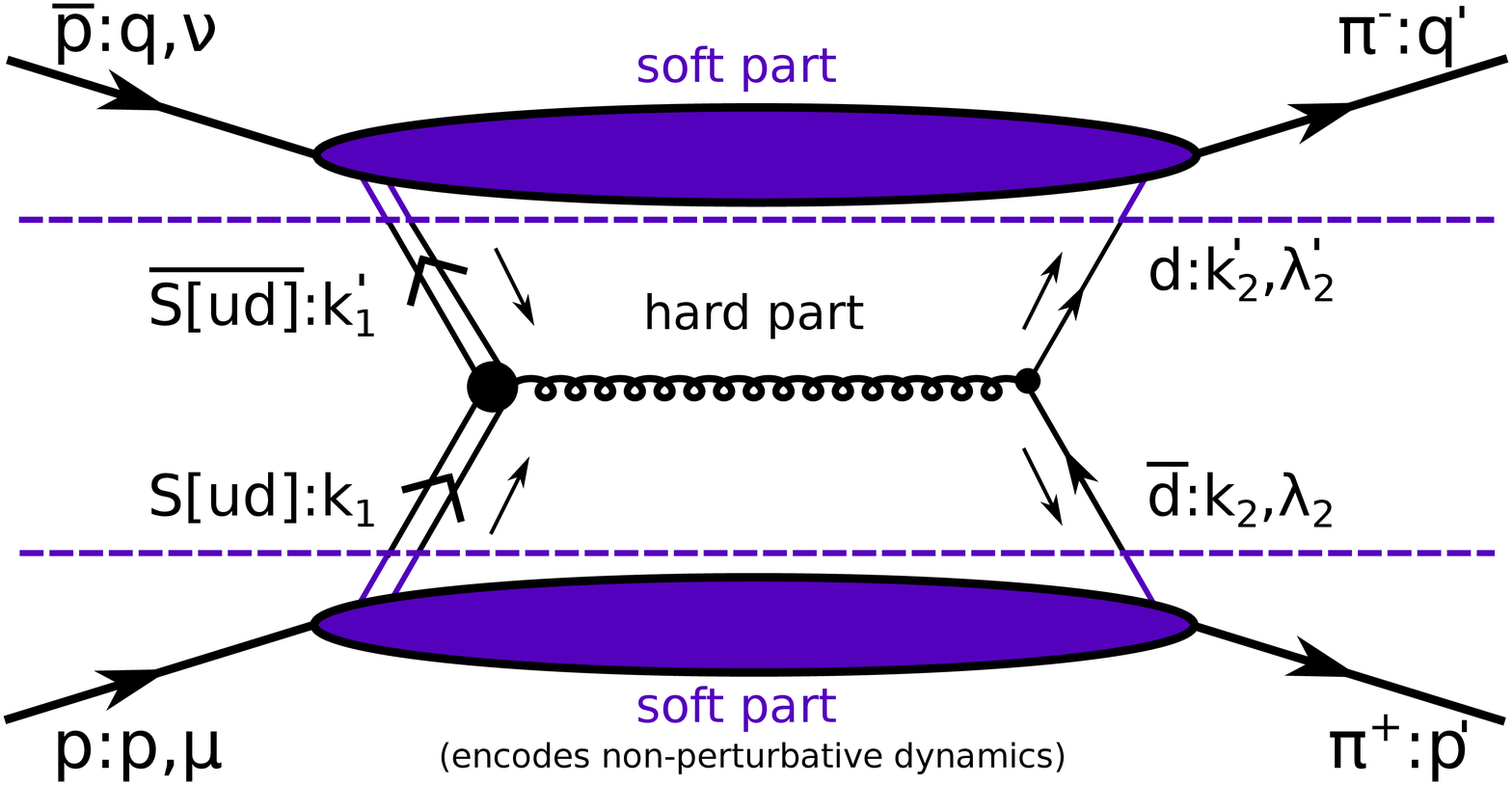}
\hfill
\includegraphics[width=.45\textwidth]{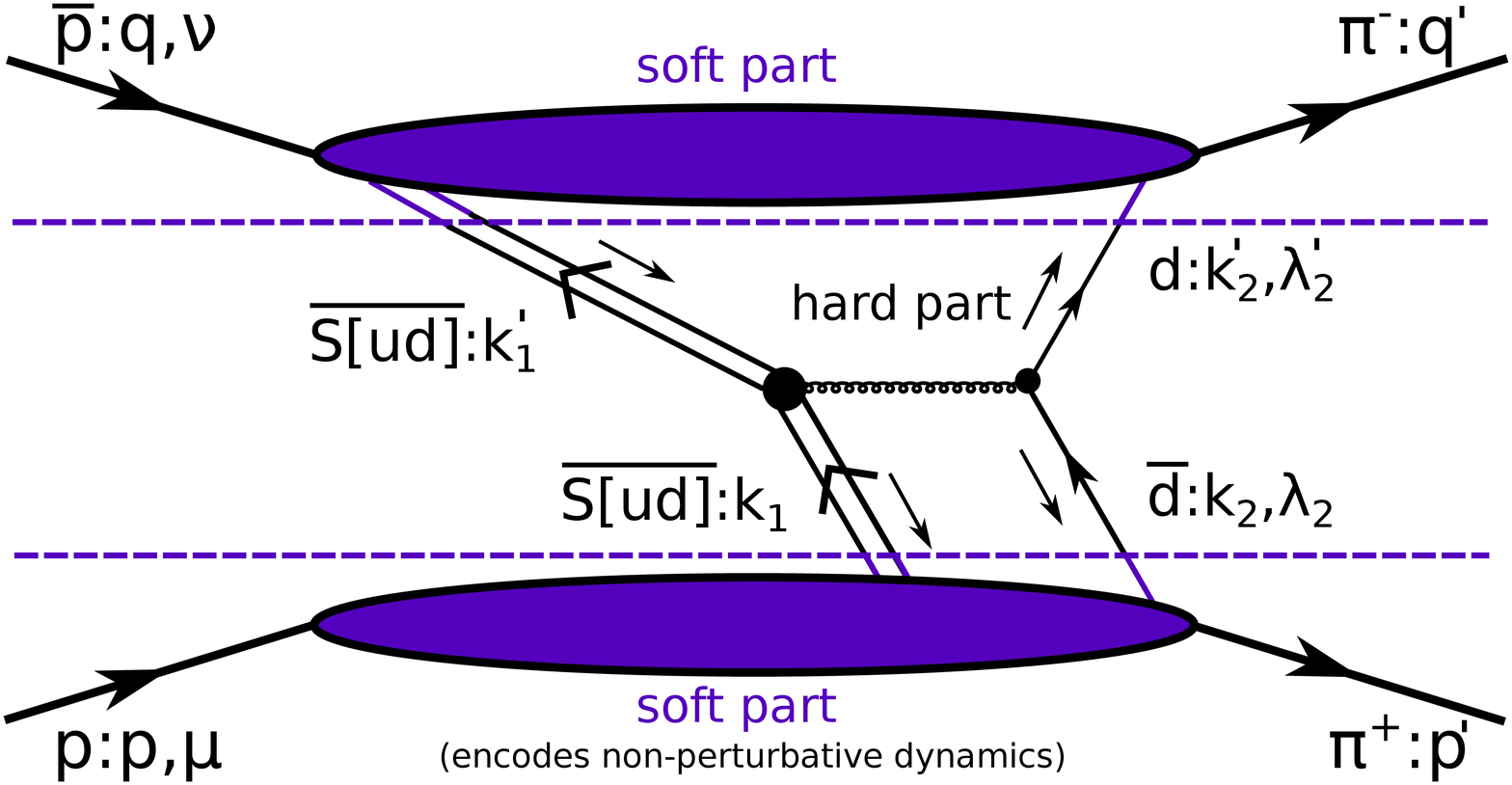}
\caption{The double-handbag mechanism for \( \process \). Left: DGLAP-contribution. Right: ERBL-contribution (one of two). The blobs contain the non-perturbative dynamics of the process and are parameterized by transition distribution amplitudes.}
\label{fig_handbag}
\end{figure}

The assignments of the momenta of the incoming baryons and outgoing mesons can be read off from Fig.~\ref{fig_handbag}.
The reference frame (symmetric center-of-momentum system (CMS)) is aligned along the \( z \)-axis of the \( 3 \)-vector of the average momentum \( \bar{\mathbf{p}} \) defined by \( \bar{p} \equiv \frac{1}{2}(p + p^{\prime})\).
Symmetric CMS means that the transverse component of the momentum transfer, defined by \( \Delta \equiv p^\prime - p = q - q^\prime \), is symmetrically shared between the incoming and outgoing particles.
In light-cone (LC) coordinates the incoming proton and the outgoing \( \pip \) momenta are written as
\begin{equation}
\label{eq_mom}
 p= \left[ \left( 1+ \xi \right)\bar{p}^+, \frac{M^2 + \mathbf{\Delta}_\perp^{2} /4}{2 \left(1 + \xi \right) \bar{p}^+}, - \frac{\mathbf{\Delta}_\perp}{2} \right] \hspace{0.5cm} \mbox{and} \hspace{0.5cm}
 p^\prime = \left[ \left( 1 - \xi \right)\bar{p}^+, \frac{m^2 + \mathbf{\Delta}_\perp^{2} /4}{2 \left(1 - \xi \right) \bar{p}^+},  \frac{\mathbf{\Delta}_\perp}{2} \right].
\end{equation}

\noindent The, so called, ``skewness parameter'' \( \xi \) is defined by \( \xi \equiv -\frac{\Delta^+}{2 \bar{p}^+} \) and gives the relative momentum transfer between the proton and the \( \pip \) in the longitudinal plus direction\footnote{Note that in our reference frame and for the pion production \( \xi < 0  \).}.
The antiproton and the \( \pim \) momenta are parameterized in an analogous way.

Similar to wide-angle Compton scattering, cf. Ref.~\cite{kroll}, we argue that for large CMS scattering angles, i.e. large Mandelstam \( s,~ t \) and \( u \) (\( t,u \ge 1 \gev \)), the process can be described by a double-handbag mechanism.
This double-handbag mechanism is shown in Fig.~\ref{fig_handbag}.
Given the large scattering angles, that provides us with a large energy scale, it is assumed that the \( \process \) amplitude factorizes into a hard-scattering kernel on the partonic level and soft hadronic \( p \rightarrow \pip \) and \( \bar{p} \rightarrow \pim  \) transition matrix elements.
We stress that there is no rigorous proof for factorization for our process.
It is, at most, a first step in this direction.
In the handbag mechanism only the minimal number of constituents which are required to transform the (initial) \( p \bar{p} \)- to the (final) \( \pip \pim \)-pair actively take part in the partonic subprocess.
Since we consider the proton to be a bound state of a scalar diquark, denoted by \( S[ud] \), and a \( u \)-quark the active constituents are the \( S[ud] \)-diquarks and \( d  \)-quarks on the proton side and the corresponding antiparticles on the antiproton side.

Using the physically plausible assumptions of restricted parton virtualities and restricted intrinsic transverse momenta, cf. Refs.~\cite{gor1,kroll,gor2}, the hadronic amplitude can be written as
\begin{align}
\begin{aligned}
 \label{eq_hadronic_amp}
 \mathscr{M}_{\mu \nu} & =
 \int \mathrm{d} \bar{x}~  \int \mathrm{d} \bar{x}^{\prime} ~ \, \widetilde{H} \left( \bar{x} ,\bar{x}^{\prime} \right) \\ 
    &\times  \bar{p}^{+}  \int \frac{\mathrm{d} z^{-}}{2\pi} e^{i  \bar{x} \bar{p}^{+}  z^{-}}  \braket{\pip:p^{\prime}|\Psi^{d} \left( -\frac{z^{-}}{2} \right) \phi^{S[ud]} \left( \frac{z^{-}}{2} \right) ~|p: p, \mu} \\ 
      & \times  \bar{q}^{-} \int \frac{\mathrm{d} z^{\prime +}}{2\pi}  e^{i \bar{x}^{\prime} \bar{q}^{-}  z^{\prime +}} \braket{\pim:q^{\prime}| ~\phi^{S[ud] \dagger} \left( \frac{z^{\prime+}}{2} \right) \bar{\Psi}^{d} \left( -\frac{z^{\prime+}}{2} \right) |\bar{p}:q,\nu}.
\end{aligned}
\end{align}
\noindent It is a convolution of a hard-scattering kernel \( \widetilde{H} \) and two hadronic matrix elements with respect to the average momentum fraction defined by
\begin{equation}
 \bar{x}^{(\prime)} \equiv \frac{\bar{k}^{ + (\prime -) }}{\bar{p}^{+}(\bar{q}^-)} \hspace{0.5cm} \mbox{with} \hspace{0.5cm} \bar{k}^{+ (\prime -)}=\frac{k_{1}^{+ (\prime - )} + k_{2}^{+ (\prime -)}}{2}.
\end{equation}

\noindent Since we are working in the LC-quantization framework we want to work with the independent degrees of freedom of the \(d\)-quark field operators.
This can be done using the same projection techniques as in Refs.~\cite{gor1,kroll,gor2}.
The amplitude thus becomes
\begin{align}
\begin{aligned}
 \label{eq_hadronic_amp2}
 \mathscr{M}_{\mu \nu} & =
 \frac{1}{4 \bar{p}^+ \bar{q}^-} \sum_{\lambda_1,\lambda_2 }\int \mathrm{d} \bar{x}~  \int \mathrm{d} \bar{x}^{\prime} ~ \, H_{\lambda_2,\lambda_2^\prime} \left( \bar{x} ,\bar{x}^{\prime} \right) \frac{1}{\bar{x} - \xi} \, \frac{1}{\bar{x}^{\prime} - \xi}\\ 
    &\times   \bar{v}\left( k_2, \lambda_2 \right) \gamma_+ \,  \bar{p}^{+} \, \int \frac{\mathrm{d} z^{-}}{2\pi} e^{i  \bar{x} \bar{p}^{+}  z^{-}}  \braket{\pip:p^{\prime}|\Psi^{d} \left( -\frac{z^{-}}{2} \right) \phi^{S[ud]} \left( \frac{z^{-}}{2} \right) ~|p: p, \mu} \\ 
      & \times  \bar{q}^{-} \int \frac{\mathrm{d} z^{\prime +}}{2\pi}  e^{i \bar{x}^{\prime} \bar{q}^{-}  z^{\prime +}} \braket{\pim:q^{\prime}| ~\phi^{S[ud] \dagger} \left( \frac{z^{\prime+}}{2} \right) \bar{\Psi}^{d} \left( -\frac{z^{\prime+}}{2} \right) |\bar{p}:q,\nu} \, \gamma^- u \left(k_2^\prime,\lambda_2^\prime \right),
\end{aligned}
\end{align}
\noindent where we have defined the hard-scattering amplitude as \( H_{\lambda_2,\lambda_2^\prime} \left( \bar{x},\bar{x}^\prime \right) \equiv \bar{u} \left( k_2^\prime , \lambda_2^\prime \right) \widetilde{H} \left( \bar{x} \bar{p}^+,\bar{x}^\prime \bar{p}^+ \right)  \)
\( \times  v \left(k_2, \lambda_2 \right)\).

\noindent Now, depending on the \( \bar{x}^{(\prime)} \)-region, three different partonic subprocesses contribute (see \\
Fig. \ref{fig_handbag}):  \(  S[ud] \, \overline{S[ud]} \rightarrow d \, \bar{d} \) (left), \( \overline{S[ud]} \rightarrow \overline{S[ud]} \,  d \, \bar{d} \) (right) and \( S[ud] \rightarrow S[ud] \, d \, \bar{d}  \) (not shown).
%
So far we have only taken a closer look on the contribution coming from the DGLAP-region alone, i.e. Fig.~\ref{fig_handbag} left. In Sec.~\ref{sec_dis} we discuss the possible influence of the ERBL-Region.

\section{Modeling the hadronic transitions}\label{sec_model}
\begin{figure}[h]
\centering
\includegraphics[width=.45\textwidth]{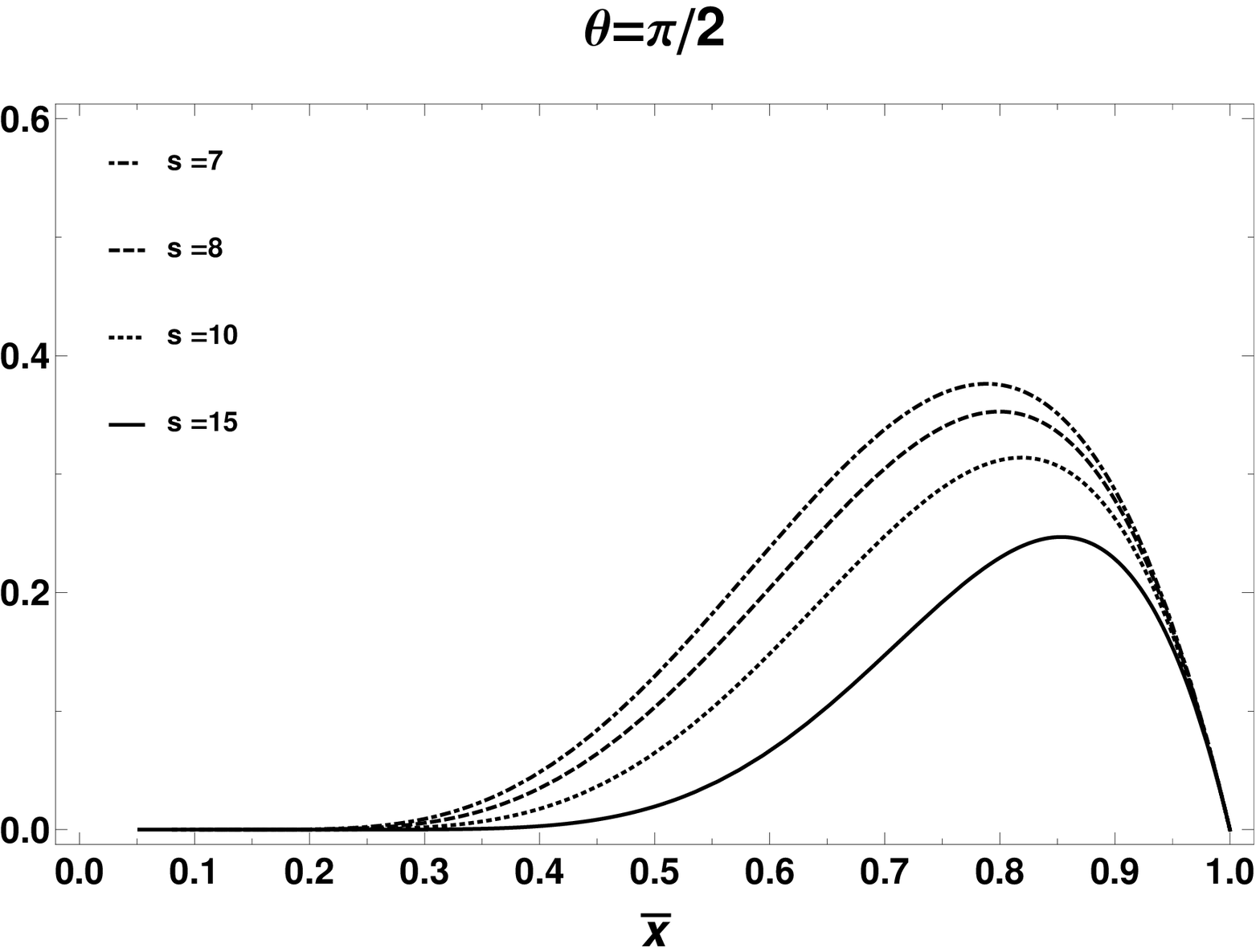}
\hfill
\includegraphics[width=.45\textwidth]{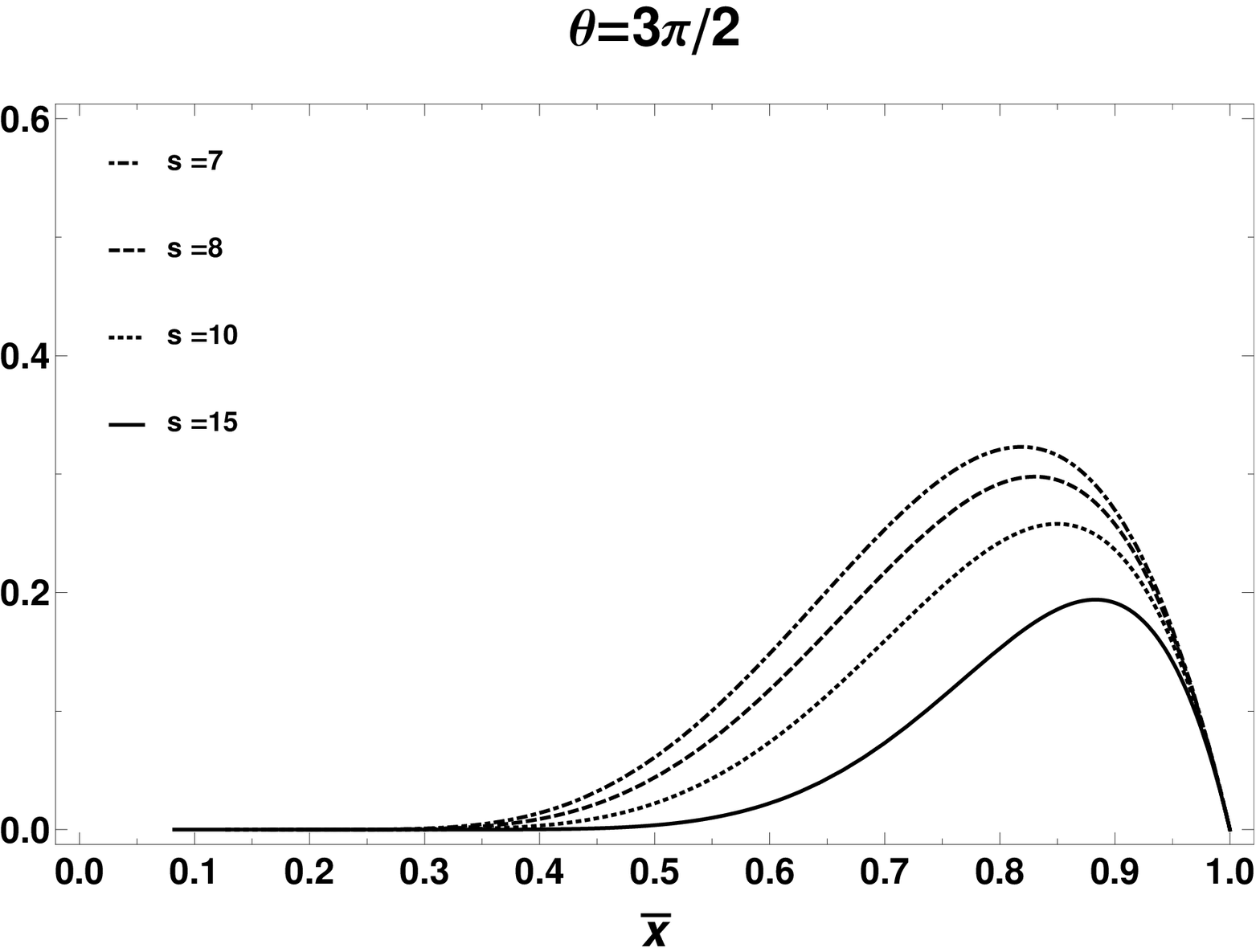}
\caption{The wave function overlap versus the average momentum \( \bar{x} \) for CMS scattering angles \( \theta = \pi/2\) (left panel) and \( \theta = 3\pi/2 \) (right panel) and different values of Mandelstam \( s\).}
\label{fig_LCWF_overlap}
\end{figure}

The soft hadronic matrix elements are modeled in the spirit of an overlap of LCWFs (focusing on the \( p \rightarrow \pip \)-transition, since the \( \bar{p} \rightarrow \pim \)-transition can be treated along the same lines).
For doing that we take the Fourier representation of the field operators and the Fock-state decomposition of the hadron states in LC-quantum field theory.
We only take the valence Fock states of the hadrons into account, i.e. a \( \ket{p:S[ud],d} \) and a  \( \ket{\pip: u,\bar{d} } \) state for the proton and \( \pip \), respectively.
Furthermore we only consider configurations with zero orbital angular momenta between the partons such that the parton helicities sum up to their parent-hadron helicity.
According to Ref.~\cite{kroll3} we choose for the proton
\begin{equation}
 \label{proton_LCWF}
 \Psi_p =  N_p \;  \tilde{x} \; \exp\left[- \frac{ a_p^{2}}{\tilde{x}(1-\tilde{x})} \, \tilde{\mathbf{k}}_{\perp}^2 \right]
\end{equation}
\noindent and according to Ref. \cite{kroll4} for the pion
\begin{equation}
 \label{pion_LCWF}
  \Psi_\pi = N_\pi \exp\left[- \frac{  a_\pi^{2}}{\hat{x}(1-\hat{x})} \, \hat{\mathbf{k}}_{\perp}^2 \right]
\end{equation}
\noindent as LCWFs.
\( \tilde{x} \) and \( \hat{x} \) denote the momentum fractions of the active \( S[ud] \)-diquark inside the proton and the active \( \bar{d} \)-quark inside the \( \pip \), respectively.
Each of the wave functions has two free parameters, a global normalization constant \( N_{p,\pi} \) and an oscillator parameter \( a_{p,\pi} \).
For the proton they are fixed by requiring a valence Fock state probability of \( P_p = 0.5 \) and a r.m.s of the intrinsic transverse momentum of the active quark of \( \sqrt{\braket{\mathbf{k}_\perp}^2} = 0.280 ~ \mbox{GeV} \).
This yields \( N_p = 61.8~\mbox{GeV}^{-2} \) and \(a_p = 1.1~\mbox{GeV}^{-1} \).
The pion parameters are fixed by demanding the r.m.s. of the active quark to be \( \sqrt{\braket{\mathbf{k}_\perp}^2} = 0.350 ~ \mbox{GeV} \) and the pion decay constant \( f_\pi = 0.093~ \mbox{GeV}\).
This gives \( N_\pi = 26.1 ~\mbox{GeV}^{-2} \) and \( a_\pi= 1.0 ~\mbox{GeV}^{-1}\).
The result for the LCWF overlap is shown in Fig.~\ref{fig_LCWF_overlap} for different values of the CMS scattering angle.
\vspace{-0.15cm}

\vspace{-0.4cm}
\section{Hard-Scattering Amplitudes}\label{sec_hard}

Under the assumptions on the parton momenta discussed in Sec.~\ref{sec_handbag} one can show that in the hard-scattering amplitude the momentum fractions are approximately \( 1\), i.e. \( x_{1,2} = x_{1,2}^{\prime} \approx 1 \). 
Since we treat the scalar diquarks and the \(d \)-quarks as being massless, the hard-scattering amplitude vanishes when \( \lambda_2 = \lambda_2^\prime \).
The non-vanishing contributions are
\begin{equation}
 \label{eq_hard_amp}
 H_{+,-} = H_{-,+}= - 4 \pi \frac{4}{9} \alpha_s (s) F_{s}(s) \sin \theta,
\end{equation}
\noindent where \( F_s(s) \), taken from Ref.~\cite{schweig}, takes care of the composite nature of the diquarks and the fact that they should dissolve into quarks if a large amount of momentum is transfered to the diquarks.

\section{Results and Discussions}\label{sec_dis}
\begin{figure}[h]
\includegraphics[width=.45\textwidth]{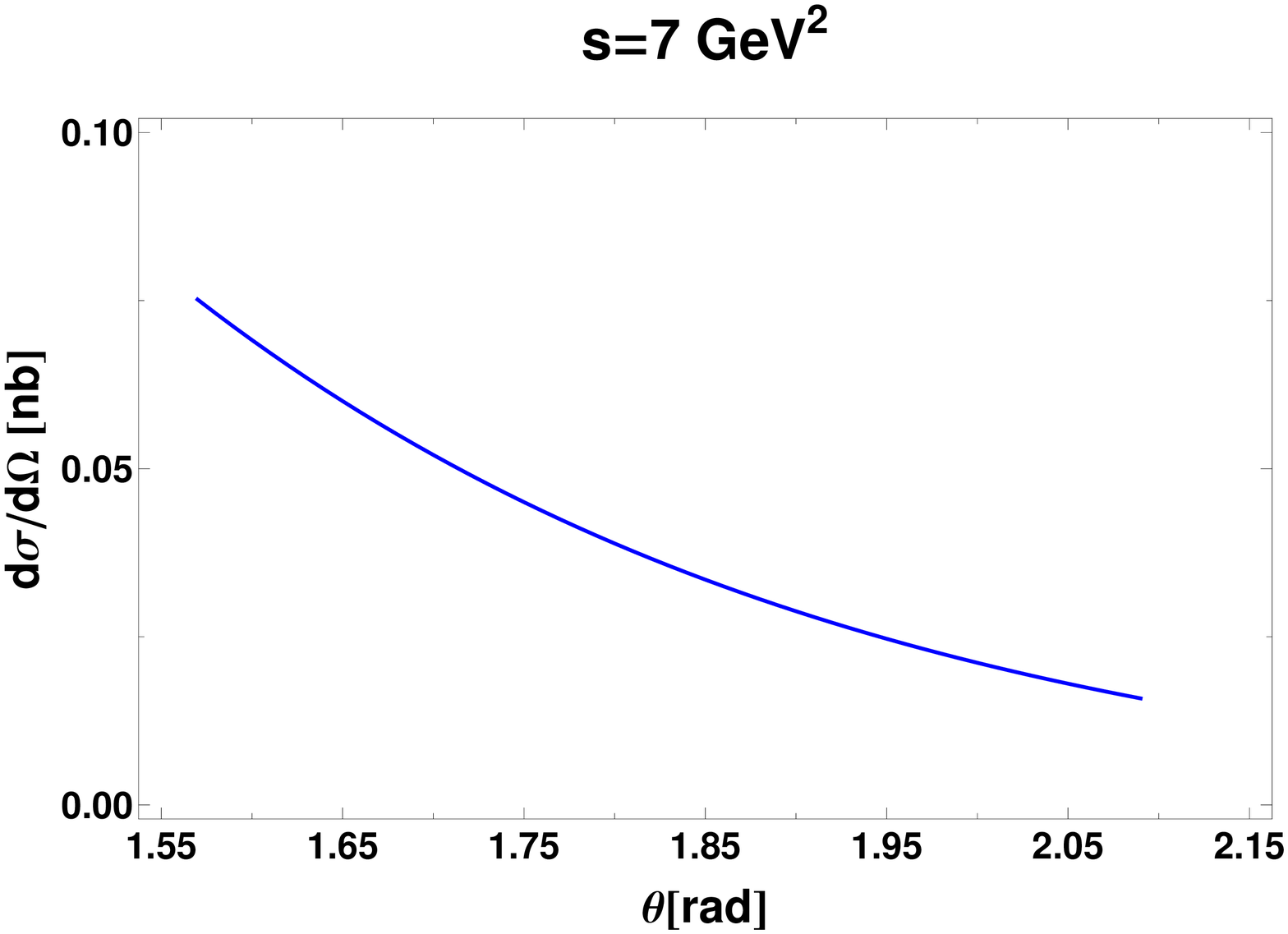}
\hfill
\includegraphics[width=.45\textwidth]{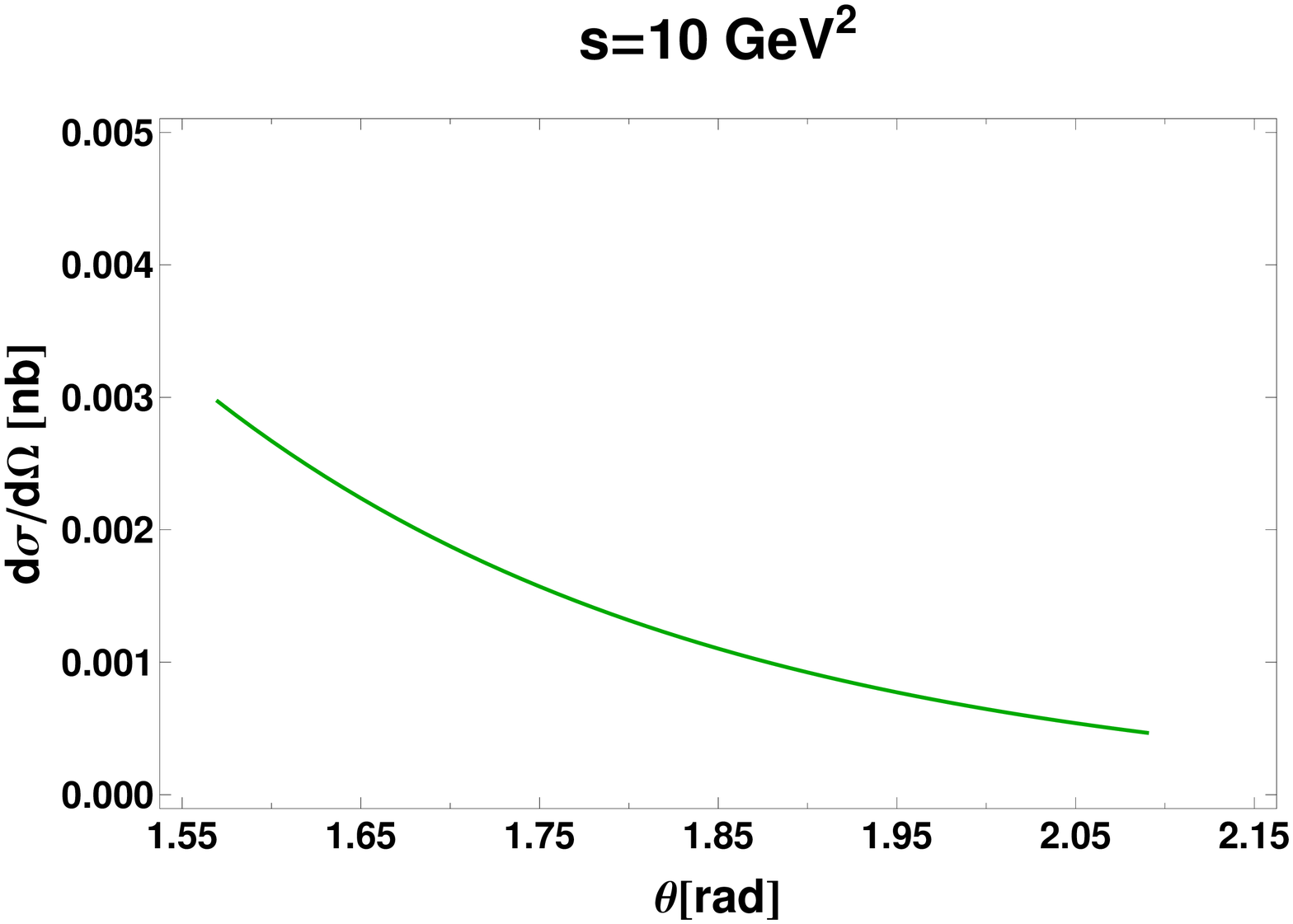}
\caption{The differential cross section for \( s = 7 \gev~ \mbox{and}~ 10 \gev \) as a function of the CMS scattering angle \( \theta \).}
 \label{fig_diff_cs}
\end{figure}

The differential cross section for \( \process \) reads
\begin{equation}
 \label{eq_cross_sec}
 \frac{d \sigma}{d \Omega} = \frac{1}{64 \pi^2} \, \frac{1}{s} \, \sqrt{\frac{1-4m^2/s}{1-4M^2/s}} \, \sigma_0 \hspace{0.5cm} \mbox{with} \hspace{0.5cm} \sigma_0 \equiv \frac{1}{4} \sum_{\mu,\nu} |\mathscr{M}_{\mu \nu}|^2 .
\end{equation}
\noindent The result for the differential cross section in the CMS-angle interval  \( \frac{\pi}{2} \le \theta \le \frac{2 \pi}{3} \) is shown in Fig.~\ref{fig_diff_cs}.
Comparing our result at \( s = 12.9 \gev \) for a CMS scattering angle of \( \pi/2 \) with the experimental data of Ref.~\cite{buran} reveals that our calculations in which we have only taken the DGLAP-region into account, predict a much smaller differential cross section. It is about three orders of magnitude smaller.

\noindent For CMS energies around \( 10 \gev \) (and larger) one may think of the ERBL-region to play a minor role, since there the skewness parameter is small\footnote{An increase of the CMS scattering angle \( \theta \) increases the skewness in magnitude in our kinematics.}, i.e. it is about \( 0.1 \) for a CMS angle of  \( \theta = \frac{2\pi}{3} \) and most of the \( \bar{x}^{(\prime)}\) integration region \( 0 \le \bar{x}^{(\prime)}  \le 1 \) is covered by the DGLAP-domain (\( \xi \le \bar{x}^{(\prime)} \le 1\)).
If this would be the case, the handbag mechanism, in its present form, would fail to give a reasonable description for \( \process \).
However, such a conclusion could be premature.
In Ref. \cite{tda2}  it is argued that the DGLAP-region dominates the cross section for large values of \( \xi \approx 1 \), while for small (and intermediate) values of \( \xi \) the dominant contribution is due to the ERBL-region, which is mainly determined by a baryon pole contribution.
Therefore it might be that the big discrepancy between our results and the experimental data can be attributed to the missing contributions from the ERBL-region. 
In order to draw definite conclusions we thus have to include also the ERBL-region.
This is the subject of current investigations. 

\ack
We want to thank the organizers of the ``Fairness 2013'' for providing an interesting workshop.
ATG and SK are supported by the Fonds zur F\"orderung der wissenschaftlichen Forschung in \"Osterreich via Grand No. J3163-N16 and FWF DK W1203-N16, respectively.
We further acknowledge helpful discussions with Bernard Pire from the  \'{E}cole Polytechnique.


\section*{References}
  \medskip
 \begin{thebibliography}{}
  \bibitem{gor1} Goritschnig A, Pire B and Schweiger W 2013 {\it Phys. Rev. D}  {\bf 87} 014017 
    \bibitem{tda1} Pire B, Semenov-Tian-Shansky and Szymanowski L 2010 {\it Phys. Rev. D } {\bf 82} 094030
  \bibitem{tda2} Pire B, Semenov-Tian-Shansky and Szymanowski L 2011 {\it Phys. Rev. D } {\bf 84} 074014
    \bibitem{tda3} Lansberg J P, Pire B, Semenov-Tian-Shansky  and Szymanowski L 2012 {\it Phys. Rev. D } {\bf 86} 114033
  \bibitem{tda4} Pire B, Semenov-Tian-Shansky and Szymanowski L 2013 {\it Phys. Lett. B } {\bf 724} 99
    \bibitem{kroll2} Diehl M, Feldmann T, Jakob R and Kroll P 2001 {\it Nucl. Phys. B} {\bf 596} 33
      \bibitem{kroll} Diehl M, Feldmann T, Jakob R and Kroll P 1998 {\it Eur. Phys. J.} {\bf C8} 409 
    \bibitem{gor2} Goritschnig A, Kroll P and Schweiger W 2009 {\it Eur. Phys. J.}  {\bf A42} 43 
  \bibitem{kroll3} Kroll P, Quadder B and Schweiger W 1989 {\it Nucl. Phys. B} {\bf 316} 373 
  \bibitem{kroll4} Feldmann T and Kroll P 2000 {\it Eur. Phys. J.} {\bf C12} 99
  \bibitem{schweig} Kroll P, Pilsner T, Sch\"urmann M and Schweiger W (1993) {\it Phys. Lett. B} {\bf 316} 546
  \bibitem{buran} Buran T et al. 1976 {\it Nucl. Phys. B} {\bf 116 } 51
 \end{thebibliography}
\end{document}